\documentclass[a4paper,11pt]{article}
\usepackage{verbatim}
\usepackage{jinstpub} 
\usepackage[utf8]{inputenc}
\usepackage{caption}
\usepackage{graphicx}
\usepackage{hyperref}
\usepackage{lineno}
\notoc

\title{\boldmath Design and performance of the prototype of the new Particle Identification system for the MAGNEX spectrometer within the NUMEN project}

\author[a,1]{C. Lombardo,\note{Corresponding author.}}
\author[a]{D. Carbone,}
\author[a]{A. Spatafora,}
\author[a]{G.A. Brischetto,}
\author[c]{D. Calvo,}
\author[a,b]{F. Cappuzzello,}
\author[a]{M. Cavallaro,}
\author[d,e]{M. Del Fabbro,}
\author[d,e]{M. Mazzocco,}
\author[d,e]{G. Montagnoli,}
\author[a,b]{R. Persiani,}
\author[c]{D. Sartirana,}
\author[a,b]{O. Sgouros,}
\author[a]{V. Soukeras,}
\author[f]{A.M. Stefanini}

\collaboration[c]{on behalf of NUMEN collaboration}

\affiliation[a]{Istituto Nazionale di Fisica Nucleare, Laboratori Nazionali del Sud, Catania, Italy}
\affiliation[b]{Dipartimento di Fisica e Astronomia “Ettore Majorana”, Università di Catania, Catania, Italy}
\affiliation[c]{Istituto Nazionale di Fisica Nucleare, Sezione di Torino, Italy}
\affiliation[d]{Dipartimento di Fisica e Astronomia “Galileo Galilei”, Università di Padova, Padova, Italy}
\affiliation[e]{Istituto Nazionale di Fisica Nucleare, Sezione di Padova, Italy}
\affiliation[f]{Istituto Nazionale di Fisica Nucleare, Laboratori Nazionali di Legnaro, Legnaro, Italy}

\emailAdd{lombardoc@lns.infn.it}

\abstract{The MAGNEX large acceptance magnetic spectrometer is installed at the Laboratori Nazionali del Sud of the Istituto Nazionale di Fisica Nucleare (INFN-LNS) in Catania.
In the past, MAGNEX experimental campaigns were performed using ion beams up to $10^{10}$ pps provided by the Superconducting Cyclotron (SC) accelerator. Nevertheless, in the context of the NUMEN (Nuclear Matrix Elements for Neutrinoless double beta decay) project, which aims to get accurate values of the tiny Double Charge Exchange (DCE) cross-sections for a large number of nuclei of interest for the neutrinoless double beta decay (0$\nu\beta\beta$), the SC is being fully refurbished featuring ion beams with energies from 15 up to 70 MeV/u and intensities up to $10^{13}$ pps. The high rate of incident particles demands a complete upgrade of the MAGNEX detectors, which will consist of a new Particle Identification (PID) system and a new gas tracker detector along with a $\gamma$-calorimeter. The PID system is equipped with a large number (720) of Silicon Carbide-Cesium Iodide (Tallium doped) telescopes. A prototype of the PID wall has been built and studied thanks to an experimental test with an $^{18}$O beam at 275 MeV incident energy performed at INFN-Laboratori Nazionali di Legnaro, Italy. In this manuscript, the design of the new PID system, the in-beam test experimental set-up, and the prototype's performances are presented.}

\keywords{Ion identification systems, Radiation-hard detectors, Focal plane detector, SiC detectors}

\begin{document}
\maketitle
\flushbottom

\section{Introduction}
\label{sec:intro}
Modern nuclear physics studies often require joining the advantages of magnetic spectrometry (such as strong rejection factors of reaction products, zero-degree measurements, etc.) with the possibility to measure heavy ions. An example is the MAGNEX large-acceptance magnetic spectrometer~\cite{cappuzzello2016magnex, cavallaro2020magnex} installed at the Laboratori Nazionali del Sud of the Istituto Nazionale di Fisica Nucleare (INFN-LNS) in Catania. It is composed by a vertically focusing quadrupole magnet and a horizontally dispersing and focusing bending dipole magnet. It ensures a large acceptance both in momentum ($\approx$ 24\%) and in solid angle (50 msr).
Developing large-acceptance magnetic spectrometers requires more advanced Focal Plane Detectors (FPD). These have to provide an unambiguous particle identification and an accurate three-dimensional tracking of the ions' trajectory downstream of the magnetic elements~\cite{fulbright1985focal, torresi2021upgraded}.
In the context of the NUMEN (Nuclear Matrix Elements for Neutrinoless double beta decay) project~\cite{cappuzzello2021numen, agodi2021numen}, which aims to get accurate values of tiny Double Charge Exchange (DCE) cross-sections for a large number of nuclei of interest for the neutrinoless double beta decay (0$\nu\beta\beta$), the Superconducting Cyclotron (SC) installed at INFN-LNS is being fully refurbished to provide ion beams with intensities up to $10^{13}$ pps.
The high rate of incident particles demands a complete upgrade of the MAGNEX FPD, which will consist of a Particle Identification (PID) system and a gas tracker~\cite{ciraldo2023characterization}. Also a new gamma calorimeter will be coupled to the spectrometer~\cite{gandolfo2023response}. In this paper, the PID system design and performances are presented. 
\section{The particle identification system}
\label{sec:pid}
\begin{figure}[htbp]
\centering
\includegraphics[width=.5\textwidth]{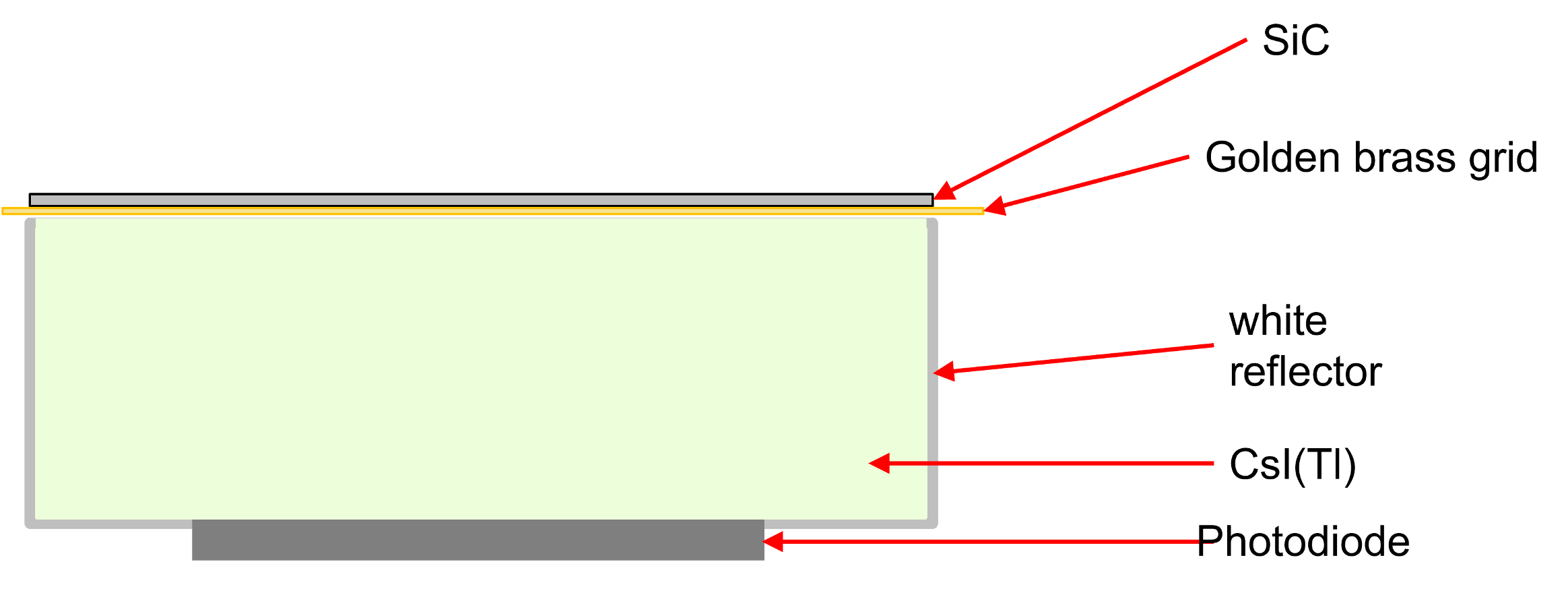}
\caption{Lateral sketch of the $\Delta$E-E telescope consisted in a SiC detector coupled with a CsI(Tl) crystal. On the bottom of the crystal, a photodiode collects the light produced inside the crystal.}
\label{fig:telescope}
\end{figure}

The NUMEN experiments require the PID system fulfillseveral requirements~\cite{cappuzzello2021numen}. Some of them are listed in the following: i) unambiguous ion identification in the region of O, F, and Ne atomic species, ii) high radiation hardness, since the expected overall heavy-ion fluence will be of the order of $10^{11}$ ions/(cm$ ^2 \cdot $ yr) in a 120-days full-power irradiation, iii) time resolution better than 2-3 ns. Moreover, it has to work in a low-pressure gas environment (tens of mbar).
To fulfill all of these requirements, a telescope solution based on thin silicon carbide detectors (SiC) coupled to inorganic scintillators Cesium Iodide-Tallium doped (CsI(Ta)) has been chosen to implement $\Delta$E-E based particle identification.
Each telescope covers an area of 1.5 cm x 1.5 cm with a dead space of 0.4 mm between each cell for a total active area of 1.54 cm x 1.54 cm.
The $\Delta$E SiC stage is 110 $\mu$m thick, featuring a 100 $\mu$m epitaxial layer plus 10 $\mu$m dead layer on the back~\cite{cappuzzello2021numen}.
Extensive studies and characterizations have been recently performed on the newly developed SiC detectors~\cite{carbone2024characterization, spataforaVIC25}. From these studies, an energy resolution of $\approx$ 0.5\% (FWHM) was deduced, which complies with the NUMEN requirements.
A 5 mm-thick 1.5 cm x 1.5 cm CsI(Tl) crystal is used to measure the residual energy. To collect the light produced by the scitillation processes, a 1 cm x 1 cm Hamamatsu photodiode (S3590-0887) is optically coupled to the crystal. 
A sketch of a single telescope is shown in figure \ref{fig:telescope}.
To cover the full focal plane of the MAGNEX spectrometer, the telescopes are arranged in 36 towers, each of them having 10 rows and 2 columns. The readout PCB board houses the pin diodes on top of which CsI(Tl) crystals are mounted. A golden brass grid is placed on the top of the crystals. 
This is used as mechanical support and ground contact for SiC detectors. SiC are wire bonded, with wires of about 18 mm length, to a second board connected to the readout PCB board.
According to the final NUMEN electronics, the signals from both SiC and CsI detectors are preamplified via the V1429 64-channels CAEN devices~\cite{Caen-preamp} and acquired via the VX2745 CAEN digitizers~\cite{Caen-digit}.
\section{In-beam test of the PID prototype}
\label{sec:exp-LNL}
\begin{figure}[htbp]
\centering
\includegraphics[width=.4\textwidth]{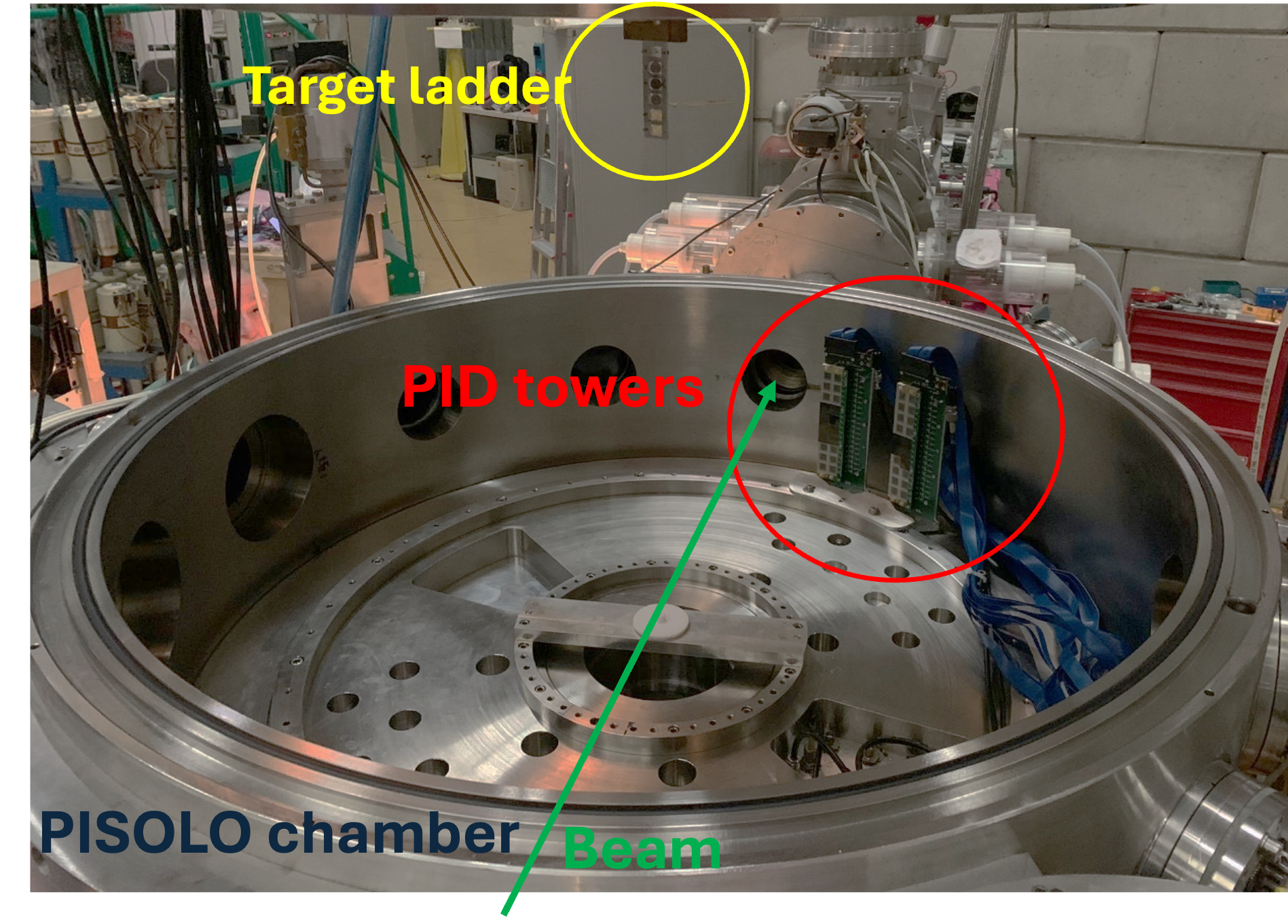}
\qquad
\includegraphics[width=.27\textwidth]{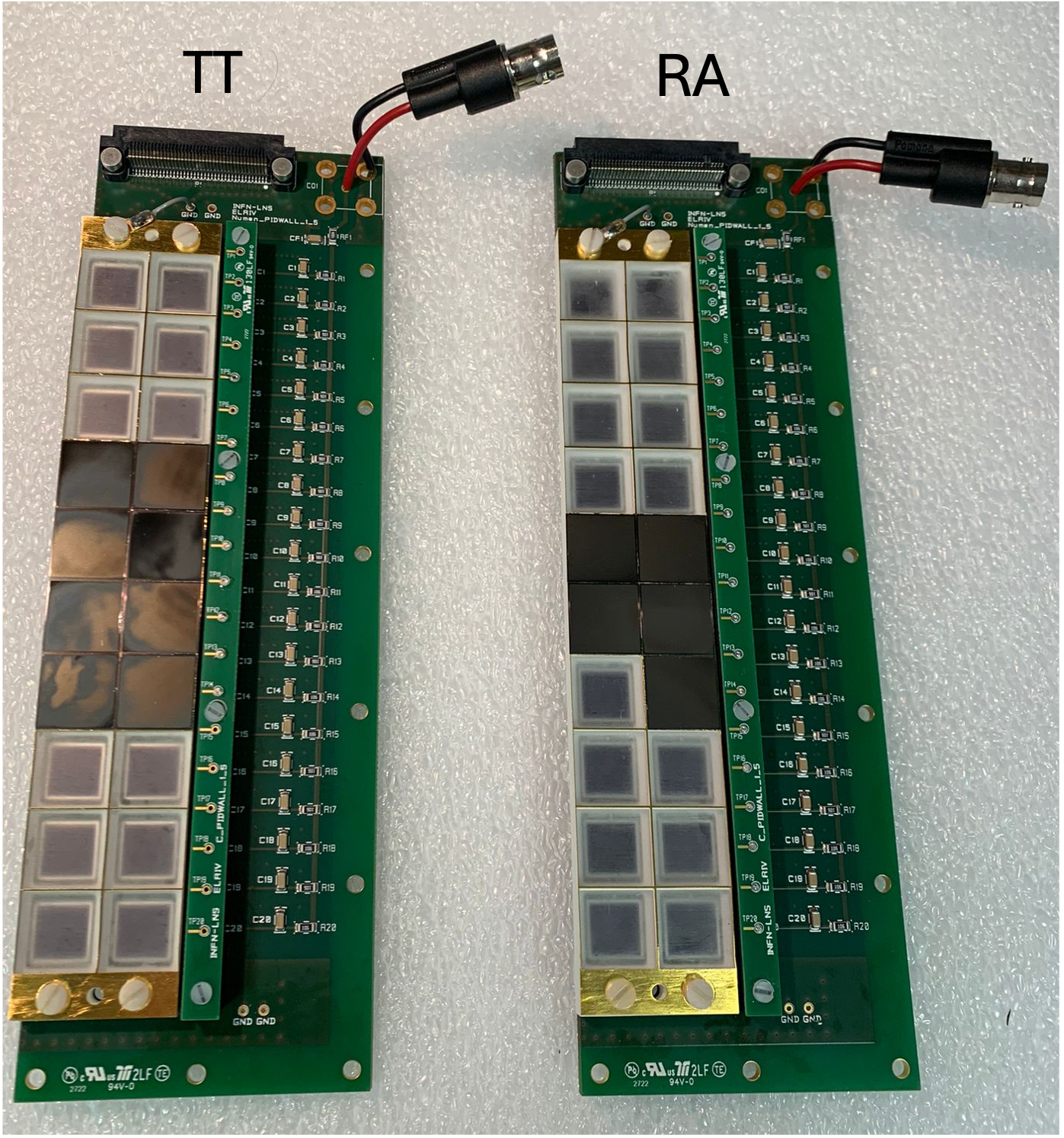}
\caption{\textit{On the left:} A picture of the PISOLO chamber at INFN-LNL. \textit{On the right:} A picture of the two PID towers.}
\label{fig:Pisolo_Towers}
\end{figure}

A new prototype of the PID wall has been recently developed, consisted in two towers. In this paper, the towers are named \textit{TT} and \textit{RA} referring to the TT0011-12 and RA0089-27 wafers from which the SiC detectors mounted in each tower were produced~\cite{carbone2024characterization}.
As shown in figure~\ref{fig:Pisolo_Towers}, the \textit{TT tower} has 8 telescopes mounted, while the \textit{RA tower} has only 5; the other tower cells hosted only the CsI(Tl) stage. The SiCs in the TT and RA towers were powered at 845 V and 352 V respectively. The photodiodes, on the other hand, had a supply voltage of 70 V.
The two towers were tested at INFN-Laboratori Nazionali di Legnaro with a $^{18}$O beam at 275 MeV incident energy impinging on a target of $^{nat}$C or $^{197}$Au.
The two towers were placed at $\sim$ 8\textdegree and at $\sim$ 40 cm downstream the target. The position of the two towers inside the PISOLO chambers is shown in figure~\ref{fig:Pisolo_Towers} alongside a picture of them.
For this test, the same front-end and read-out electronics that will be used in NUMEN was adopted, as described in section~\ref{sec:pid}.\\ 
The first step of the data reduction is to construct the coincidence spectrum between the SiC and the CsI(Tl) signals in order to choose a proper time window.  We found that a 300 ns time window, opened 1.9 $\mu$s after the SiC signal, is an optimized choice to select the coincidence peak between the two stages of the telescopes.
This coincidence condition was adopted in all the further steps of the data analysis which results are presented in the following section.
\section{Results}
\label{sec:results}
The main goals of this experiment are the evaluation of the Z identification capabilities, the validation of the NUMEN readout and DAQ for the PID system, and the DCE cross-section sensitivity estimation.
The results obtained are promising and summarized in the following.
\subsection{Atomic number resolution}
\label{subsec:z}
\begin{figure}[htbp]
\centering
\includegraphics[width=.45\textwidth]{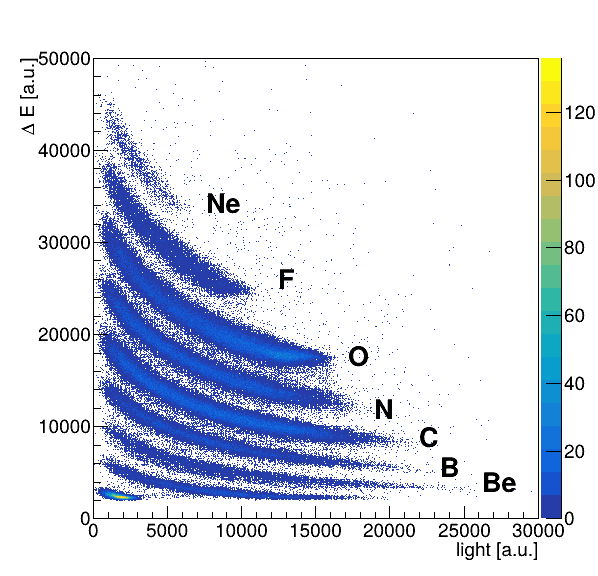}
\qquad
\includegraphics[width=.45\textwidth]{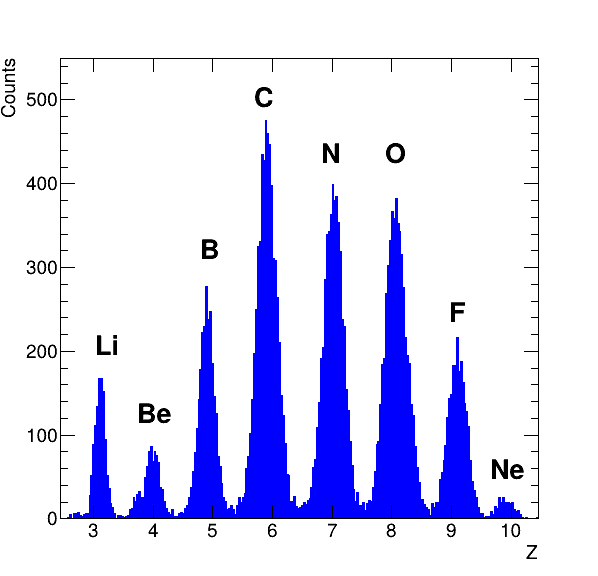}
\caption{\textit{On the left:} $\Delta$E-E matrix with the PID system. \textit{On the right:} Occurrence of different ions identified by the PID system. Both plots are obtained with the $^{18}$O beam@15 AMeV with the carbon target.}
\label{fig:Z_res_and DEE}
\end{figure}

When the $^{18}$O beam interacts with the C target, various nuclear reactions occur, causing the production of different ions that can be intercepted and identified by the PID system. An example of a $\Delta$E-E matrix is shown on the left panel of figure~\ref{fig:Z_res_and DEE}.
The separation between different light ion species is clear. In order to extract the resolution in Z, the histogram of Z-species occurrence was produced and is shown on the right panel of figure~\ref{fig:Z_res_and DEE}. In the Ne region, we obtained $\frac{\Delta Z}{Z}\sim$ 3.3\% $\simeq \frac{1}{30}$ better than the value of $\frac{1}{24}$ previosly quoted with a prototipal single telescope, as reported in ref.~\cite{cappuzzello2021numen}, and much better than the value of $\frac{1}{10}$ requested by the NUMEN experiments to clearly identify the O, F and Ne ions.
\subsection{Cross-section sensitivity}
\label{subsec:cross-section}
\begin{figure}[htbp]
\centering
\includegraphics[width=.65\textwidth]{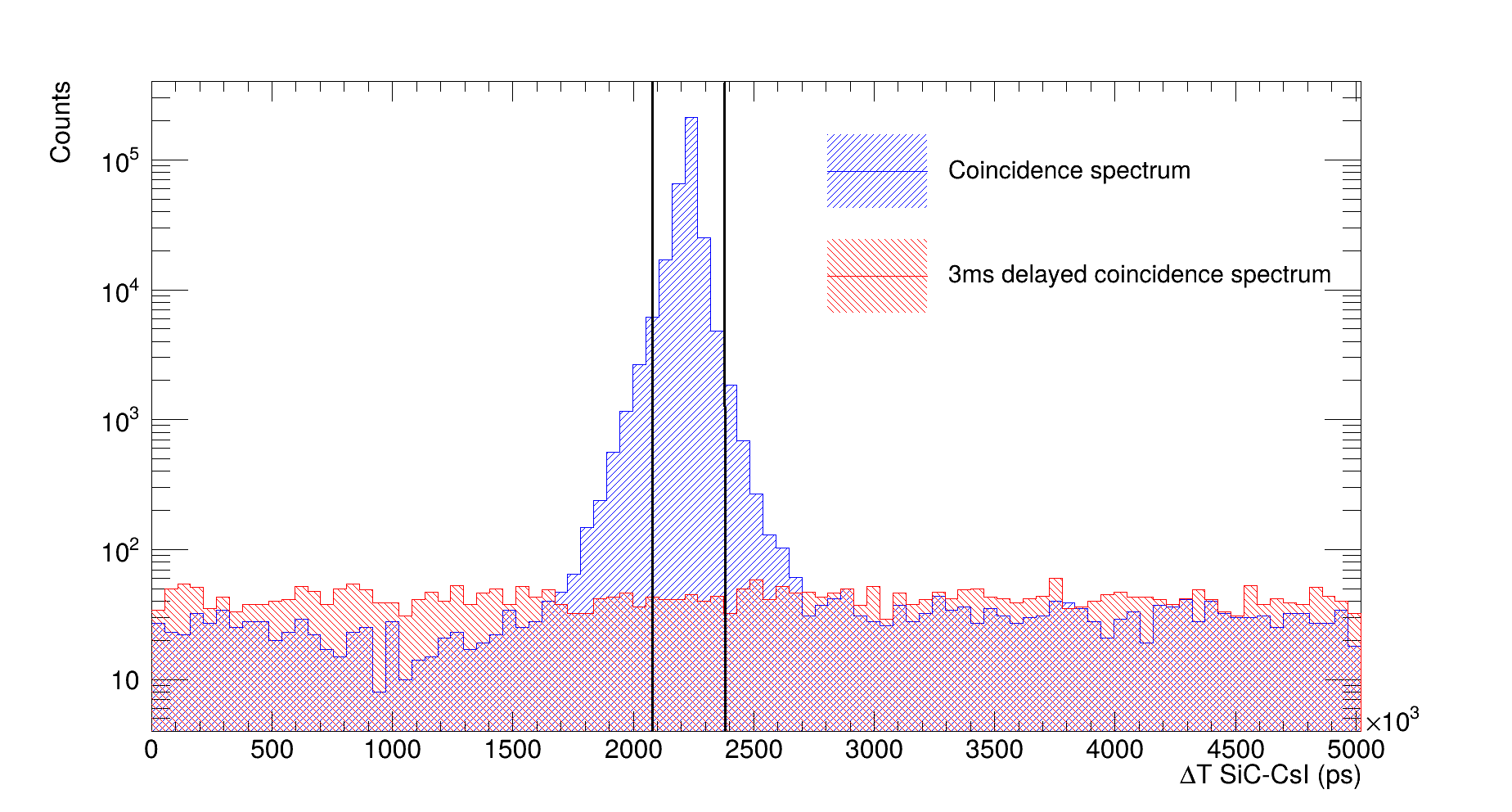}
\caption{Time distribution of the events with the coincidence gate opened after the SiC with (red) and without (blue) the 3 ms delay. The black vertical lines highlight the 300 ns coincidence gate.}
\label{fig:coincidence}
\end{figure}

The NUMEN project aims to measure the cross-section of the ground state to ground state (g.s. to g.s.) DCE reactions down to few nb for isotopes candidate for neutrinoless double beta decay~\cite{soukeras2021measurement, eke2024measurement}. Thus, MAGNEX detectors must have a low background level to ensure the significance in the measurement of such tiny cross-section values.
Part of the background events comes from the spurious coincidence events happening in the telescopes of the PID system. In figure~\ref{fig:Z_res_and DEE}, a typical SiC-CsI coincidence spectrum is reported for one telescope of the TT tower and the data acquired with the carbon target. The grey band in figure~\ref{fig:Z_res_and DEE} represents the 300 ns coincidence gate, chosen as the optimal compromise between signal collection and background rejection. The number of spurious events beneath the coincidence peak, was evaluated opening the the coincidence gate 3 ms after the SiC signal, where only spourios coincidence are expected to occur, and is shown by the red spectrum of figure~\ref{fig:Z_res_and DEE}.
Figure~\ref{fig:coincidence} shows the time distribution of the coincidence events with the coincidence gate (5 $\mu s$ long) opened after the SiC with and without the 3 ms delay.
The resulting average amount of spurious coincidences in the Neon region is 0.4\%.
From this measurement, it was possible to estimate the cross-section sensitivity over all the $\Delta Z_{proj} = +2$ charge-changing reactions, applying the Rutherford cross-section normalization factor extracted from the study of the elastic scattering data on the Au target, which were recorded in the same experimental conditions. We found an average charge-changing cross section sensitivity of $\approx$ 2.6 nb for the \textit{TT tower} and $\approx$ 1.5 nb for the \textit{RA} one for all the reactions involving Ne ejectiles.
Since NUMEN is interested in cross-section sensitivity on g.s. to g.s. transition in the DCE transition channel, we estimated the spurious backgound in the $^{18}$Ne region introducing the <<MAGNEX-like>> selection conditions, related to the full PID technique~\cite{cavallaro2020magnex} and to the MAGNEX energy resolution ($\Delta$E/E = 1/1000), corresponding to a total reduction factor of $\approx$ 0.01.
Taking into account this reduction factor, the cross-section sensitivity in the g.s. to g.s. DCE channel is better than 0.03 nb and 0.02 nb in the \textit{TT} and \textit{RA towers}, respectively.
These values are better than the cross-section sensitivity for the DCE g.s. to g.s. region of 0.30 $\pm$ 0.03 nb obtained with the previous MAGNEX FPD~\cite{calabrese2020analysis} and are compliant with the NUMEN requirements.

\section{Conclusion}
\label{sec:concl}
The Superconducting Cyclotron at INFN-LNS is being fully refurbished, featuring ion beams with energies from 15 up to 70 MeV/u and intensities up to 10$^{13}$ pps. The high rate of incident particles requires a complete upgrade of the MAGNEX FPD. The new PID system has been designed and is composed by 720 telescopes of SiC detectors coupled with CsI(Tl) crystals. The telescopes are arranged in 36 towers, of 20 telescopes each.
With SiC detectors produced from two wafers, two towers were constructed and tested with an $^{18}$O beam at 275 MeV incident energy impinging on C and Au targets.
The results are promising, since the system is capable of identifying ions in our region of interest (C, O, F, Ne) with a resolution $\Delta Z/Z\simeq$ 3.3\% in the Ne region.
Moreover, the estimated cross-section sensitivity in the g.s. to g.s. region of DCE reactions is better than 0.03 nb for the \textit{TT tower} and better than 0.02 nb for the \textit{RA} one. 
Other sources of background are expected from partial charge collection at the borders of the telescopes. Geant4 simulations are ongoing to study this source of background.
The prototipal towers will also be tested in a low-pressure gas environment to study the performances and the coupling with the future gas-tracker detector.
\acknowledgments
This project received funding from the European Union ‘‘Next Generation EU’’ (PNRR M4 - C2 – Inv. 1.1 - DD n. 104 del 02-02-2022 - PRIN 20227Z4HB8)


\bibliographystyle{JHEP}
\bibliography{biblio2.bib}

\end{document}